\documentclass[12pt,preprint]{aastex}

\begin{document}
\newcommand{\etal}{{\it et al.}}
\include{idl}

\title{Precision X-ray Timing of RX J0806.3+1527 with CHANDRA:
Evidence for Gravitational Radiation from an Ultracompact Binary}
\author{Tod E. Strohmayer} \affil{Laboratory for High Energy
Astrophysics, NASA's Goddard Space Flight Center, Greenbelt, MD 20771;
stroh@clarence.gsfc.nasa.gov}

\begin{abstract}

RX J0806.3+1527 is a candidate double degenerate binary with possibly
the shortest known orbital period. The source shows an $\approx 100
\%$ X-ray intensity modulation at the putative orbital frequency of
3.11 mHz (321.5 s). If the system is a detached, ultracompact binary
gravitational radiation should drive spin-up with a magnitude of
$\dot\nu \sim 10^{-16}$ Hz s$^{-1}$. Efforts to constrain the X-ray
frequency evolution to date have met with mixed success, principally
due to the sparseness of earlier observations.  Here we describe the
results of the first phase coherent X-ray monitoring campaign on RX
J0806.3+1527 with {\it Chandra}. We obtained a total of 70 ksec of
exposure in 6 epochs logarithmically spaced over 320 days.  With these
data we conclusively show that the X-ray frequency is increasing at a
rate of $3.77 \pm 0.8 \times 10^{-16}$ Hz s$^{-1}$. Using the
ephemeris derived from the new data we are able to phase up all the
earlier {\it Chandra} and ROSAT data and show they are consistent with
a constant $\dot\nu = 3.63 \pm 0.06 \times 10^{-16}$ Hz s$^{-1}$ over
the past decade. This value appears consistent with that recently
derived by Israel et al. largely from monitoring of the optical
modulation, and is in rough agreement with the solutions reported
initially by Hakala et al., based on ground-based optical
observations.  The large and stable $\dot\nu$ over a decade is
consistent with gravitational radiation losses driving the
evolution. An intermediate polar (IP) scenario where the observed
X-ray period is the spin period of an accreting white dwarf appears
less tenable because the observed $\dot\nu$ requires an $\dot m
\approx 2 \times 10^{-8}$ $M_{\odot}$ yr$^{-1}$, that is much larger
than that inferred from the observed X-ray luminosity (although this
depends on the uncertain distance and bolometric corrections), and it
is difficult to drive such a high $\dot m$ in a binary system with
parameters consistent with all the multiwavelength data.  If the
ultracompact scenario is correct, then the X-ray flux cannot be
powered by stable accretion which would drive the components apart,
suggesting a new type of energy source (perhaps electromagnetic) may
power the X-ray flux.

\end{abstract}


\section{Introduction}

The influence of gravitational radiation emission on binary evolution
is most easily perceived in very close binaries. Double degenerate
systems containing a pair of white dwarfs are the most compact systems
known and can theoretically have orbital periods shorter than 5
minutes (Warner 1995; Tutukov \& Yungelson 1996; Nelemans et
al. 2001). Two candidate double degenerate systems have been the focus
of much research over the last few years; V407 Vul (also known as RX
J1914.4+2456), and RX J0806.3+1527 (hereafter J0806). Both objects
were discovered in the ROSAT survey (Beuermann et al. 1999), and they
share much in common (see Cropper et al. 2003 for a recent review).
Each shows a periodic, $\approx 100 \%$ X-ray modulation with a sharp
rise and more gradual decline. The observed X-ray periods are 9.5 and
5.4 minutes for V407 Vul and J0806, respectively. They are also
optically variable at these same periods, and the optical lightcurves
lag their X-ray profiles by about 1/2 a cycle, strongly suggesting
that the optical variations result from X-ray heating of the secondary
in a phase-locked, synchronized binary (Ramsay et al. 2000; Israel et
al. 2003; Israel et al. 2004).

Although the current ``even money'' bet is that the observed periods
in these systems represent the orbital periods of ultracompact
systems, there is still no direct detection of orbital motion, as
would be provided by observations of Doppler shifted spectral lines,
for example. Scenarios associating the observed periods with the spin
of a white dwarf are not completely ruled out, but are highly
constrained. For example, recent optical and near-IR photometry of
J0806 strongly constrains the nature of its secondary. Only an
implausibly large distance would allow a low mass, main-sequence
secondary (Reinsch, Burwitz \& Schwarz 2004). This poses difficulties
for an intermediate polar (IP) interpretation for J0806 (Norton,
Haswell \& Wynn 2004). Other challenges for such models are the
extremely soft X-ray spectrum (Israel et al. 2003), and the lack of
any detection of the longer orbital period that is usually seen in
IPs.

Important constraints on models for these systems can be obtained by
measuring their frequency evolution.  If the systems are Roche lobe
accretors, and the mass donors are degenerate, then stable accretion
will lead to a widening of the orbit and a decrease in the orbital
frequency, ie. $\dot\nu < 0$ (Strohmayer 2002; Marsh, Nelemans \&
Steeghs 2004). This is contrary to what has been observed to
date. Based on a timing study of archival ROSAT and ASCA data
Strohmayer (2002) found evidence that the orbital frequency of V407
Vul is increasing at a rate consistent with loss of orbital angular
momentum to gravitational radiation. More recent monitoring of the
system with {\it Chandra} and XMM-Newton confirms the initial evidence
for spin-up (Strohmayer 2004; Ramsay et al. 2005). Hakala et
al. (2003, 2004) have used archival ROSAT and optical timing
measurements of J0806 to attempt to constrain its frequency
evolution. Using noncoherent methods they found evidence for a
frequency derivative, $\dot\nu$, of either about 3 or $6 \times
10^{-16}$ Hz s$^{1}$, but could not unambiguously measure the rate due
to the sparseness of available observations.  Strohmayer (2003)
explored whether the early ROSAT data and a single, more recent {\it
Chandra} observation were consistent with the range of $\dot\nu$
values found by Hakala et al. (2003). He concluded they were, and
found weak evidence favoring the higher $\dot\nu$ solution. More
recently, Israel et al. (2004) have recently reported results from an
optical monitoring campaign to explore the frequecy evolution of
J0806. They obtained sufficient optical coverage to perform a coherent
timing analysis and find a positive $\dot\nu = 3.5 \pm 0.1 \times
10^{-16}$ Hz s$^{-1}$. Using their coherent optical solution they were
able to show that existing X-ray observations were also consistent
with this solution.

The current timing results suggest that either accretion does not
power the X-ray flux, or perhaps the donors are non-degenerate. This
latter alternative also appears unlikely, especially for J0806, simply
because so compact a system could not contain such a donor (Savonije,
de Kool \& van den Heuvel 1986). If the systems are detached and
ultracompact then the question arises as to the source of the X-ray
emission. An interesting possibility, suggested by Wu et al. (2002),
is that these objects are ``electric'' stars, powered by unipolar
induction.  In such a scenario the primary is magnetized and a small
asynchronism between the primary and secondary induces an
electromotive force that drives currents between the components.  This
process can heat a small polar cap on the primary, thus producing
X-ray emission. Although difficulties with the details of this model
exist (see, for example, Barros et al. 2005) it has the attractive
feature that the orbital spin-up is dominated by the loss of
gravitational radiation. In light of the remaining uncertainties
regarding the nature of these objects and the interesting implications
of timing constraints, it is important to continue temporal monitoring
of these objects in both the optical and X-ray domain.

Here we report results of the first phase coherent X-ray timing
campaign for J0806 using {\it Chandra}. We obtained a total of 70 ksec
of exposure in 6 logarithmically spaced epochs spanning 320 days.  The
observing plan was designed to maintain phase coherence assuming
$\dot\nu$ was as large as $\approx 10^{-15}$ Hz s$^{-1}$. We show that
the new {\it Chandra} data conclusively establish that the X-ray
frequency of J0806 is increasing at a rate of $\approx 3.6 \times
10^{-16}$ at the present epoch.  We use our new ephemeris to phase
connect earlier {\it Chandra} and ROSAT data and show that the source
has been spinning up at a more or less constant rate over the past
decade.  We discuss the implications of the now secure conclusion that
the X-ray and optical frequencies are phase locked and are increasing
at a large rate consistent with that expected due to gravitational
radiation from a detached, ultracompact binary. We argue that spin-up
of an accreting white dwarf appears increasingly unlikely because the
large accretion rate required is inconsistent with the modest inferred
X-ray luminosity and the difficulty of driving so high a mass transfer
rate in a system consistent with all the multiwavelength constraints.

\section{Data Extraction and Analysis}

We observed J0806 with {\it Chandra} on six occasions from January 5,
2004 to November 22, 2004. We used ACIS-S in continuous clocking (CC)
mode in order to mitigate pile-up. We used the backside-illuminated
(S3) chip to maximize the soft photon response.  Table 1 contains a
summary of the observations. To prepare the data for our timing
analysis, we first corrected the detector readout times to arrival
times using the {\it Chandra} X-Ray Center's (CXC) analysis thread on
timing with CC mode data. We then corrected the arrival times to the
solar system barycenter using the CIAO tool {\it axbary} with the
JPL-DE405 ephemeris. We used the source position, $\alpha = 08^h 06^m
23^s.2$, $\delta = 15^{\circ} 27' 30''.2$, which is consistent with
both optical and X-ray observations (Ramsay, Hakala \& Cropper 2002;
Israel et al. 2002; Israel et al. 2003). This process produced a set
of photon arrival times in the barycentric dynamical time system
(TDB).

The CC mode produces a one-dimensional ``image" of the sky exposed to
the detectors. An example image for the 2004, January 5 observation is
shown in Figure 1. Photons from J0806 produce the strong peak in the
plot. We carried out the same image analysis for all observations and
extracted only events from within the source peaks for our timing
study. Figure 2 shows a portion of a light curve produced from source
extracted events and demonstrates that {\it Chandra} easily detects
individual pulses from the source. We carried out this procedure on
all the data and obtained a total of 15,418 photon events for our
timing analysis.

\subsection{Coherent Timing Solutions}

We performed a coherent timing analysis using the $Z_n^2$ statistic
(Buccheri 1983; see also Strohmayer 2004 and Strohmayer \& Markwardt
2002 for examples of the use of this statistic in a similar
context). To model the arrival times of pulses we use a two parameter
phase model, $\phi (t) = \nu_0 (t-t_0) + \frac{1}{2} \dot\nu (t -
t_0)^2$. Here, $\nu_0$ is the frequency at the reference epoch, $t_0$,
and $\dot\nu$ is a constant frequency derivative. Since details of the
method are described elsewhere we do not repeat them here.

We began by computing $Z_3^2$ as a function of $\nu_0$ assuming a
model with $\dot\nu \equiv 0$. This is more or less equivalent to a
power spectrum analysis. The results are shown in Figure 3. The best
constant frequency is 3.1101430 mHz, and there is no ambiguity with
regard to identifying the correct frequency.  That is, sidebands
caused by the gaps in the data are at vastly less significant values
of $Z_3^2$ (see Strohmayer 2004). We next computed a set of phase
residuals using this constant frequency model (see Figure 5, upper
panel). This $\dot\nu \equiv 0$ model does not fit the phases well,
indeed, the need for a positive $\dot\nu$ is indicated in this plot by
the downward opening quadratic trend in the residuals.

We next included non-zero $\dot\nu$ values in the model and performed
a grid search by calculating $\chi^2$ at each $\nu_0$ - $\dot\nu$ pair
(see Strohmayer 2004 for details of the method). The results are
summarized in Figures 4 and 5. Figure 4 shows the $68\%$ and $90\%$
confidence contours (dashed) in the $\nu_0$ - $\dot\nu$ plane. We find
a best fitting solution of $\nu_0 = 3.1101380 \pm 0.0000006$ mHz and
$\dot\nu = 3.77 \pm 0.8 \times 10^{-16}$ Hz s$^{-1}$, using a
reference epoch of $t_0 = 53009.889943753$ MJD (TDB). Errors here are
1$\sigma$. The fit is excellent, with a minimum $\chi^2 = 48.9$ with
51 degrees of freedom (dof). Fixing $\dot\nu$ at zero results in an
increase in $\chi^2$ of about 87, which strongly excludes the constant
frequency model. These calculations demonstrate conclusively that
J0806 is indeed spinning up. Figure 5 shows two set of phase
residuals. As noted above, the top panel shows the residuals obtained
from the best constant frequency ($\dot\nu = 0$) model, and the bottom
panel shows the best solution with a positive $\dot\nu$.

With an accurate solution in hand we can project backwards in time to
the epoch of the earliest {\it Chandra} and ROSAT observations of
J0806 (these data were discussed previously by Strohmayer 2003; Israel
et al. 2003; and Hakala et al. 2003).  We first included the November,
2001 {\it Chandra} data, and recomputed $\chi^2$ on our $\nu_0$ -
$\dot\nu$ grid. The results are also shown in Figure 4 (solid
contours, again, $68\%$ and $90\%$ confidence). As can be seen, the
November, 2001 {\it Chandra} data are entirely consistent with the
solution derived from our 2004 data (dashed contours), but the longer
baseline provides much tighter constraints on the
parameters. Combining all the {\it Chandra} data we find $\nu_0 =
3.11013824 \pm 0.00000017$ mHz and $\dot\nu = 3.63 \pm 0.06 \times
10^{-16}$ Hz s$^{-1}$, using the same reference epoch. This solution
is very close to that reported by Israel et al. (2004) based primarily
on their optical monitoring campaign.  There is no requirement for a
$\ddot \nu$ term, with an upper limit of $\approx 2 \times
10^{-24}$ Hz s$^{-2}$.

We show two representations of the phase residuals using all the {\it
Chandra} data. Figure 6 shows the phase residuals with respect to a
constant frequency ($\dot\nu = 0$) model, and further demonstrates
that a quadratic trend (positive $\dot\nu$) is clearly needed to model
all the phases.  Figure 7 shows the phase residuals using our best
$\nu_0$ - $\dot\nu$ model. This model accurately describes all the
{\it Chandra} phase timings, and has an rms residual of less than 0.01
of a cycle (horizontal dashed lines).  We note that this solution is
also consistent with the earliest ROSAT data, but the inclusion of
those data do not appreciably tighten the constraints because the
contributions to $\chi^2$ are dominated by the more numerous and
higher signal to noise ratio {\it Chandra} measurements. Finally, we
phase folded all the {\it Chandra} data using our best timing
solution, and the resulting modulation profile is shown in Figure 8.

\section{Discussion and Implications}

Based on the Israel et al. (2004) study and the present work, it is
now undeniable that the X-ray and optical modulations of J0806 are
stable, phase-locked and speeding up at a rate consistent with what
would be expected for gravitational radiation losses in an
ultracompact binary.  What still has not been demonstrated
conclusively is that the observed period is indeed orbital in
nature. If the period is not orbital, the only remaining plausible
model would seem to be accretion-induced spin-up of a white dwarf,
most likely in a nearly face-on IP system similar to that described by
Norton, Haswell \& Wynn (2004). Assuming this model is correct one can
estimate the accretion rate required to account for the observed
spin-up. To order of magnitude the accretion-induced spin-up rate is,
\begin{equation}
I \dot\omega = \dot m \left ( G M r_c \right )^{1/2} \; .
\end{equation}
Here, $I$, $\dot\omega= 2\pi\dot\nu$, $\dot m$, $M$, and $r_c$ are the
stellar moment of inertia, the spin angular frequency derivative, the
mass accretion rate, the stellar mass, and the characteristic radius
at which the accreted matter is ``captured'' by the star (effectively
the lever arm over which the torque acts). Assuming $I = \frac{1}{5} M
R^2$ for white dwarfs (see Marsh et al. 2004), with $R$ the stellar
radius, we can express the mass accretion rate as,
\begin{equation}
\dot m = \frac{2\pi}{5} \dot\nu \left ( \frac{M R^4}{G r_c} \right
)^{1/2} \; .
\end{equation}
Plugging in our measured $\dot\nu$, assuming $M= 0.5 M_{\odot}$, using
Eggleton's mass-radius relation (see Verbunt \& Rappaport 1988), and
taking $r_c = R$ we obtain a characteristic value of $\dot m \approx
2.6 \times 10^{-8}$ $M_{\odot}$ yr$^{-1}$. If the primary's magnetic
field channels the flow then $r_c$ can be larger than $R$. A more
characteristic value may be the circularization radius (Verbunt \&
Rappaport 1988). For a binary with an orbital period in the 1 - 3 hr
range, and a total system mass of $1 M_{\odot}$ this could increase
$r_c$ by perhaps a factor of 3, and reduce the mass accretion rate by
$\sqrt 3$. These arguments, which should be considered accurate at the
order of magnitude level, would provide an accretion luminosity,
$L_{acc} = GM\dot m / R$ in the range from $0.8 - 1.3 \times 10^{35}$
ergs s$^{-1}$, which is equivalent to a flux (at 500 pc) of $f_{acc}
\approx 2 - 4 \times 10^{-9}$ ergs cm$^{-2}$ s$^{-1}$. We note that a
reduction in mass by a factor of 2 would also reduce the accretion
luminosity by almost a factor of 2.5 (with $r_c = R$), so constraints
on the system mass are also important for an accurate understanding of
its energetics.

We are still in the process of carrying out a spectral study of our
CC-mode data (ACIS CC-mode is still only crudely calibrated for
spectroscopy), however, Israel et al. (2003) have done a spectral
analysis of the ACIS-S imaging data from November, 2001. Assuming the
source spectrum has not changed dramatically over time we can use
their spectral parameters to scale our observed count rates to X-ray
fluxes. They found a peak, unabsorbed X-ray flux in the 0.1 - 2.5 keV
band of $1.5 \times 10^{-11}$ ergs cm$^{-2}$ s$^{-1}$. We calculated
average count rates for each of our observations and show them plotted
versus time in Figure 9. Our brightest epoch had an average rate of
0.26 s$^{-1}$, and the time history shows $\approx 50\%$ variations in
intensity. For comparison, the average count rate during the November,
2001 observations reported by Israel et al. (2003) was 0.29 s$^{-1}$,
a bit higher than in our more recent observations. We note that recent
count rates could be reduced somewhat by the additional build-up of a
contaminant on the ACIS detectors (Marshall et al. 2004). We also note
that spectral results obtained recently using XMM-Newton data give
comparable flux levels (Israel et al. 2004).

The observed X-ray fluxes from J0806 are, at a minimum, several orders
of magnitude less than the total accretion luminosity expected if we
are seeing accretion-induced spin-up of a white dwarf. We note that
larger, previously measured fluxes do not falsify this argument
because we have measured the spin-up rate and fluxes over the same
epoch.  Although it is possible that a significant fraction of the
accretion luminosity appears outside the X-ray band--particularly in
the extreme UV--it may be difficult to explain all of the mismatch in
this way. As noted by Israel et al. (2003), sensitive EUV observations
of the source would be very revealing on this score. A second caveat
is the uncertain distance.  The most recent spectral study with
XMM-Newton suggests the absorption column is consistent with the
Galactic value in the direction to J0806 (Israel et al. 2004).  This
implies that unless the source is very far out of the galactic plane
its distance is probably not much greater than about 500 pc. The fact
that its proper motion is small provides some evidence that it is not
a halo object and therefore is not too far out of the Galactic plane
(Israel et al. 2002).

Another question which can be asked is whether an IP system with an
orbital period somewhat longer than one hour (as suggested by Norton
et al. 2004), and with a very late type secondary consistent with the
optical and infrared photometry can in fact transfer mass at the rate
required to match the observed spin up?  It appears likely that in
such a system angular momentum losses due to gravitational radiation
would be insufficient to account for such a large mass transfer
rate. This is because gravitational radiation driven mass loss is a
very sensitive function of orbital frequency and secondary mass
(Rappaport et al. 1982; Priedhorsky \& Verbunt 1988; Marsh, Nelemans
\& Steeghs 2004).  With an orbital period longer than 1 hr, and a
secondary later than M6 (Reinsch et al. 2004), a system with stellar
parameters appropriate for an IP interpretation of J0806 would have a
long term mass transfer rate driven by gravitational radiation of
$\dot m_{gr} \approx 5 \times 10^{-11}$ $M_{\odot}$ yr$^{-1}$, which
is much smaller than required to achieve the observed spin up rate in
J0806 unless the presently observed rate is uncharacteristic of the
long term rate. This could be the case in transient systems, for
example. Although J0806 is variable there is no indication from past
observations that it is a transient. A class of binaries with
characteristic transfer rates of a few $10^{-11}$ $M_{\odot}$
yr$^{-1}$ are the recently discovered accreting millisecond pulsars,
with orbital periods from 40 min to a few hours, and very low mass
secondaries (see Wijnands 2004 for a recent review).  The main
difference between these systems and a putative IP scenario for J0806
would seem to be that their primaries are neutron stars and not white
dwarfs.  Perhaps other mechanisms can drive mass transfer at the
required rate or for some reason we are currently observing a higher
than average transfer rate. Because of these uncertainties we realize
this argument should be considered with a bit of caution.

If the observed period is orbital then the system is a powerful source
of gravitational radiation. Indeed, a circular binary with component
masses $m_1$ and $m_2$ separated by a distance $a$ will radiate a
gravitational wave luminosity (Peters \& Matthews 1963),
\begin{equation}
L_{gw} = \frac{32}{5} \frac{G^4}{a^5 c^5} \left ( m_1^2 m_2^2 (m_1 +
m_2) \right ) \; .
\end{equation}
If the orbital decay results only from gravitational radiation losses
and there is no mass transfer, then the constraint on $\dot\nu$
implies a constraint on the so-called chirp mass
\begin{equation}
\left ( \frac{M_{\rm ch}}{M_{\odot}} \right )^{5/3} = \left (
\frac{\mu}{M_{\odot}} \right ) \left ( \frac{m_1 + m_2}{M_{\odot}}
\right )^{2/3} = 2.7 \times 10^{16} \left ( \frac{\nu}{10^{-3}\; {\rm
Hz}} \right )^{-11/3} \dot\nu \;\; ,
\end{equation}
where $\mu = m_1 m_2 / ( m_1 + m_2 )$ is the reduced mass. Assuming
equal masses we find a value $m_1 = m_2 = 0.37 M_{\odot}$ for our
measured $\dot\nu$. Using a separation, $a$, consistent with the
inferred orbital period one finds $L_{gw} \approx 2 \times 10^{35}$
ergs s$^{-1}$, which is substantially larger than the observed X-ray
flux.

\section{Summary and Conclusions}

Our {\it Chandra} monitoring of J0806 confirms that its X-ray
frequency is increasing at a rate of $3.6 \times 10^{-16}$ Hz
s$^{-1}$. Our results are in agreement with the independent optical
monitoring campaign carried out by Israel et al. (2004). These studies
show that the X-ray and optical modulations are phase-locked and
stable over more than a decade.  Although direct confirmation that the
observed period is orbital in nature is still lacking, we believe that
the present evidence favors an orbital interpretation. If this is the
case, then we are seeing the influence of gravitational radiation
losses on the most compact binary known. Indeed, the rate of change of
its orbital frequency would be $\approx 10^5$ times larger than that
of the famous binary pulsar (Taylor \& Weisberg 1989).

Because of the remaining uncertainties and the importance of this
object in the context of directly observing gravitational radiation
driven orbital evolution, and perhaps studying a new form of stellar
energy (ie. electric stars), additional observations are extremely
important. Deep phase resolved optical and X-ray spectroscopy might
reveal radial velocity variations consistent with orbital motion.
Sensitive EUV data would be helpful in constraining the overall energy
budget and thus bounding the accretion rate. Continued X-ray and
optical timing are essential to further study the torque and phase
stability. Finally, if further electromagnetic observations do not
prove definitive, then observations with a future space-based
gravitational radiation detector, such as the NASA/ESA LISA mission
currently in development, would provide a final test of the
ultracompact hypothesis.  Indeed, if the observed period is orbital,
then the source will produce a strong gravitational radiation signal
at twice the observed electromagnetic frequency, and LISA should be
able to detect it easily.  LISA observations could also provide the
distance and inclination of the system. With such information in hand,
an intimate portrait of the evolution and energetics of the system
would emerge.

\acknowledgements

It is a pleasure to acknowledge the insightful comments of the referee. 

\centerline{\bf References}

\noindent{} Barros, S.~C.~C., Marsh, T.~R., Groot, P., Nelemans, G.,
Ramsay, G., Roelofs, G., Steeghs, D., \& Wilms, J.\ 2005, MNRAS, 357,
1306.

\noindent{} Beuermann, K. et al. 1999, A\&A, 347, 47.

\noindent{} Buccheri, R. et al. 1983, A\&A, 128, 245.

\noindent{} Cropper, M., Ramsay, G., Wu, K. \& Hakala, P. 2003, in
Proc. Third Workshop on Magnetic CVs, Cape Town, (astro-ph/0302240).

\noindent{} Hakala, P., Ramsay, G. \& Byckling, K. 2004, MNRAS, 353,
453.

\noindent{} Hakala, P. et al. 2003, MNRAS, 340, L10.

\noindent{} Israel, G. L. et al. 2004, Proceedings of the XLVII
National Conference of the S.A.It., Mem.S.A.It. Suppl.,
Eds. A. Wolter, G.L. Israel \& F. Bacciotti, (astro-ph/0410453).

\noindent{} Israel, G. L. et al. 2003, apJ, 598, 492.

\noindent{} Israel, G. L. et al. 2002, A\&A, 386, L131.

\noindent{} Israel, G. L. et al. 1999, A\&A, 349, L1.

\noindent{} Marsh, T. R., Nelemans, G. \& Steeghs, D. 2004, MNRAS, 350, 113.

\noindent{} Marshall, H. L., Tennant, A., Grant, C. E., Hitchcock,
A. P., O'Dell, S. L., \& Plucinsky, P. P. 2004, Proc. SPIE, 5165, 497.

\noindent{} Nelemans, G., Portegies Zwart, S. F., Verbunt, F. \& Yungelson, 
L. R. 2001, A\&A, 368, 939.

\noindent{} Norton, A. J., Haswell, C. A. \& Wynn, G. A. 2004, A\&A,
419, 1025.

\noindent{} Peters, P.~C., \& Mathews, J.\ 1963, Physical Review , 131, 435.

\noindent{} Priedhorsky, W. C. \& Verbunt, F. 1988, ApJ, 333, 895.

\noindent{} Ramsay, G., Hakala, P., Wu, K., Cropper, M., Mason, K.~O.,
C{\' o}rdova, F.~A., \& Priedhorsky, W.\ 2005, MNRAS, 357, 49.

\noindent{} Ramsay, G., Cropper, M., Wu, K., Mason, K.~O., \& Hakala, P.\ 
2000, MNRAS, 311, 75.

\noindent{} Ramsay, G., Hakala, P. \& Cropper, M. 2002, MNRAS, 332, L7.

\noindent{} Ramsay, G., et al. 2002, MNRAS, 333, 575.

\noindent{} Rappaport, S., Joss, P. C. \& Webbink, R. F. 1982, ApJ, 254, 616.

\noindent{} Reinsch, K., Burwitz, V. \& Schwarz, R. 2004, To appear in
RevMexAA(SC) Conference Series, Proc. of IAU Colloquium 194 `Compact
Binaries in the Galaxy and Beyond', La Paz (Mexico), eds. G. Tovmassian \& 
E. Sion, (astro-ph/0402458).

\noindent{} Savonije, G. J., de Kool, M., \& van den Heuvel,
E. P. J. 1986, A\&A, 155, 51.

\noindent{} Strohmayer, T. E. 2004, ApJ, 610, 416.

\noindent{} Strohmayer, T. E. 2003, ApJ, 593, L39.

\noindent{} Strohmayer, T. E. 2002, ApJ, 581, 577.

\noindent{} Strohmayer, T. E., \& Markwardt, C. B., 2002, ApJ, 577, 337.

\noindent{} Taylor, J. H. \& Weisberg, J. M. 1989, ApJ, 345, 434.

\noindent{} Tutukov, A. \& Yungelson, L. 1996, MNRAS, 280, 1035.

\noindent{} Verbunt, F., \& Rappaport, S. 1988, ApJ, 332, 193.

\noindent{} Warner, B. W. 1995, Cataclysmic Variable Stars (Cambridge:
Cambridge Univ. Press).

\noindent{} Wijnands, R. 2004, in ``X-ray Timing 2003: Rossi and
Beyond,'' AIP Conference Proceedings vol. 714, ed. P. Kaaret,
F. K. Lamb, \& J. H. Swank, pg. 209.

\noindent{} Wu, K., Cropper, M., Ramsay, G. \& Sekiguchi, K. 2002, MNRAS, 
331, 221.

\noindent{} Zapolsky, H. S. \& Salpeter, E. E. 1969, ApJ, 158, 809.

\pagebreak

\centerline{\bf Figure Captions}

\figcaption[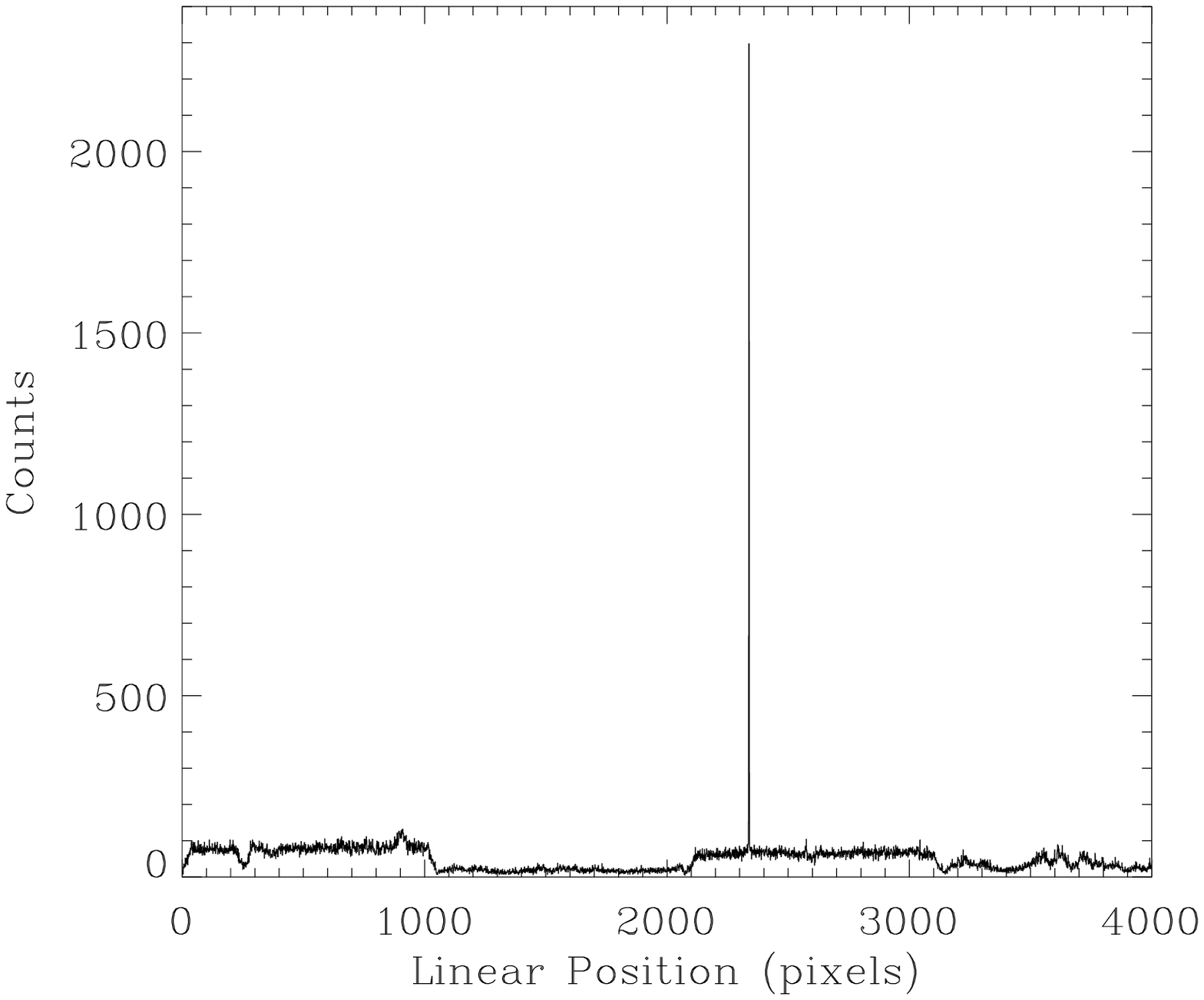]{{\it Chandra} one-dimensional image of J0806 from
our January 5, 2004 (UTC) ACIS-S CC-mode observation. The sharp peak
contains photons from the source.
\label{f1}}

\figcaption[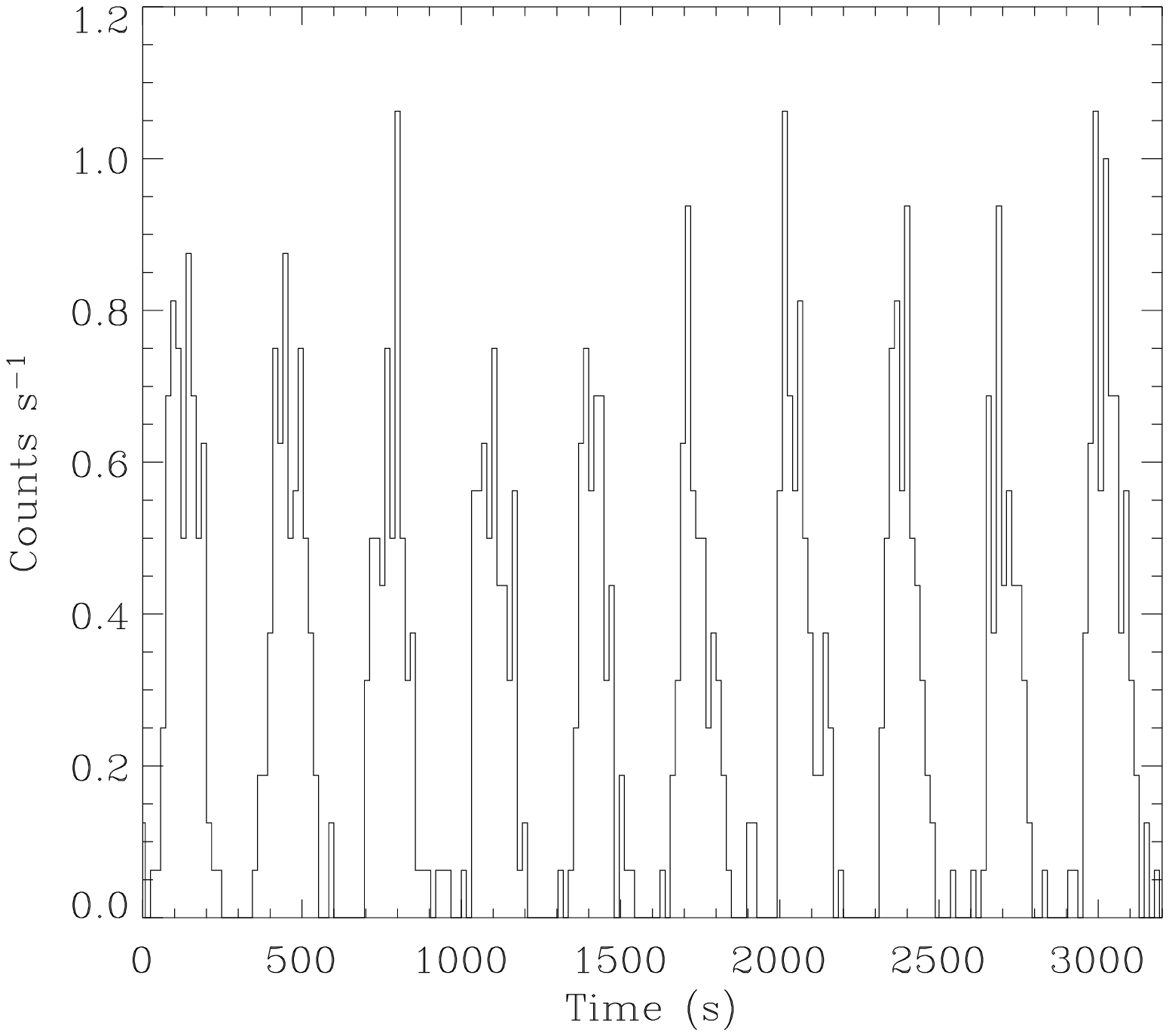]{Light curve of a portion of the {\it Chandra}
ACIS-S data from the January 5, 2004 (UTC) observation of J0806. The
bin size is 16 s, and ten individual pulses are shown.
\label{f2}}

\figcaption[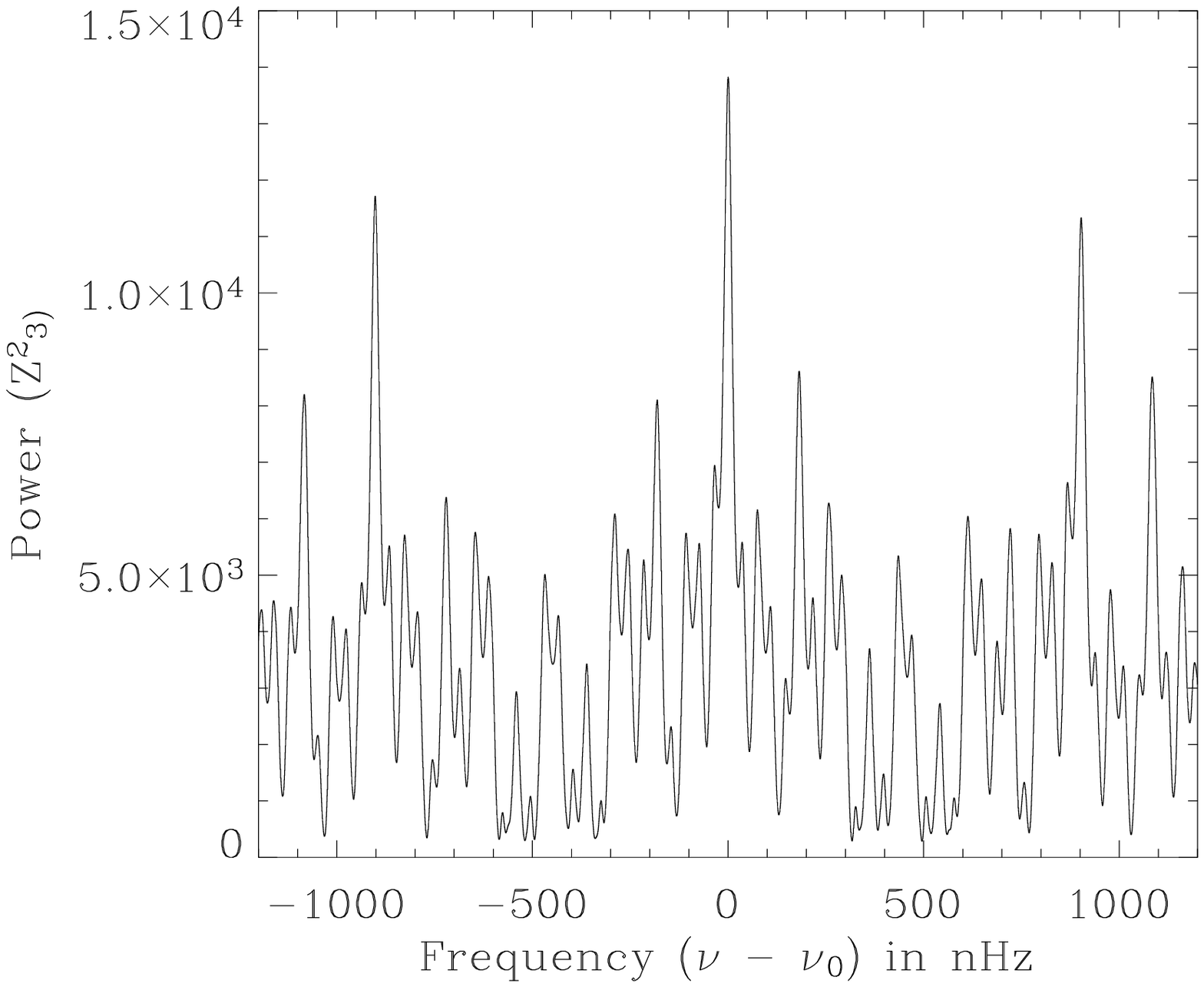]{Best constant frequency measurement for J0806
using all of our epoch 2004 {\it Chandra} observations.  The $Z_3^2$
power spectrum is shown as a function of frequency in the vicinity of
$\nu_0 = 3.11014250$ mHz.
\label{f3}}

\figcaption[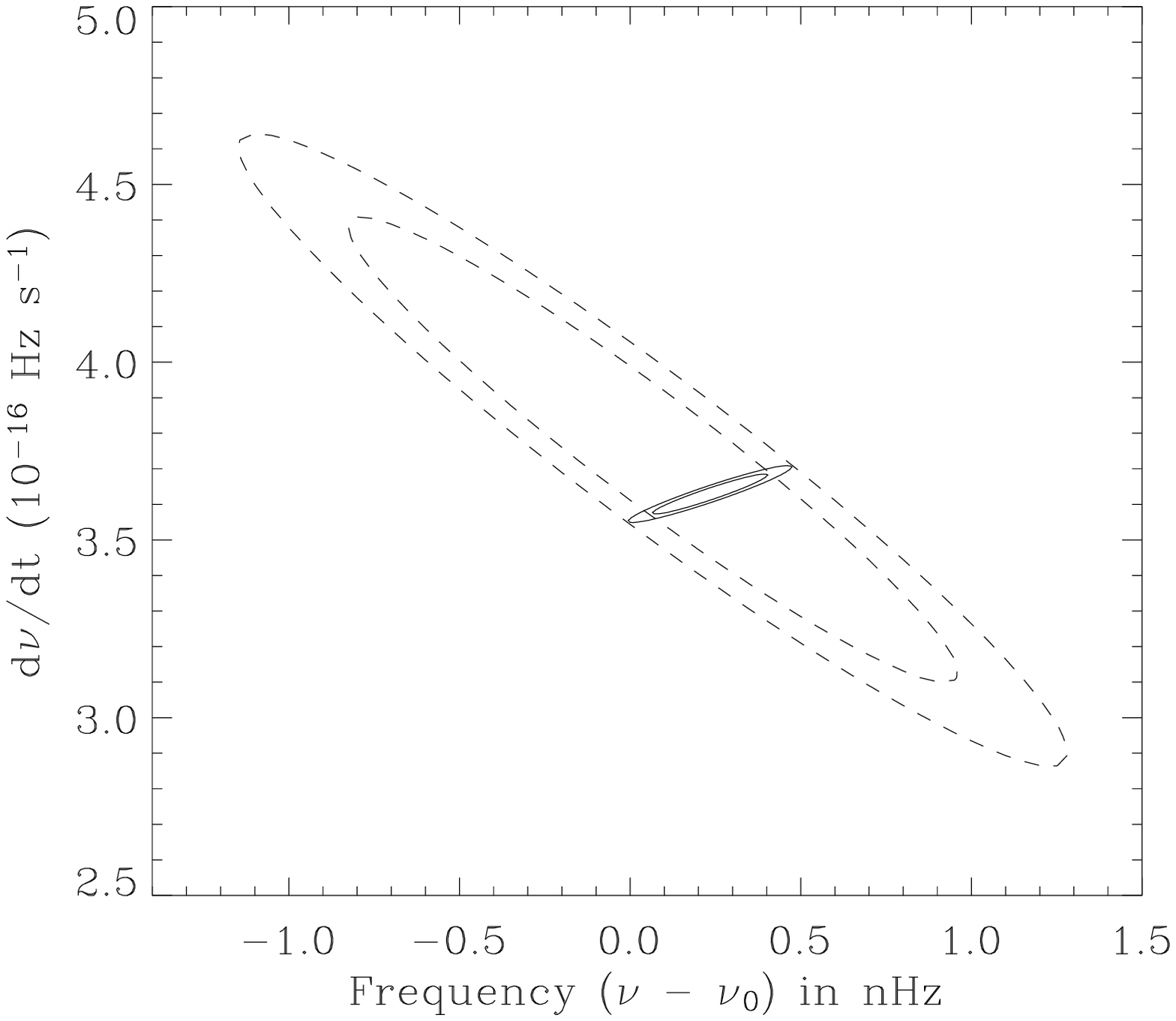]{Constraints on $\nu_0$ and $\dot\nu$ for J0806
from our phase-timing analysis using the new, epoch 2004 {\it Chandra}
dataset (dashed contours), and after combining the 2001 and 2004
datasets (solid contours). We show the joint 68\% and 90\% confidence
regions in each case.  The results definitively show that J0806 is
spinning up at at rate of $3.6 \times 10^{-16}$ Hz s$^{-1}$. Here
$\nu_0 = 3.1101380$ mHz, and the reference epoch, $t_0 =
53009.889943753$ MJD (TDB).
\label{f4}}

\figcaption[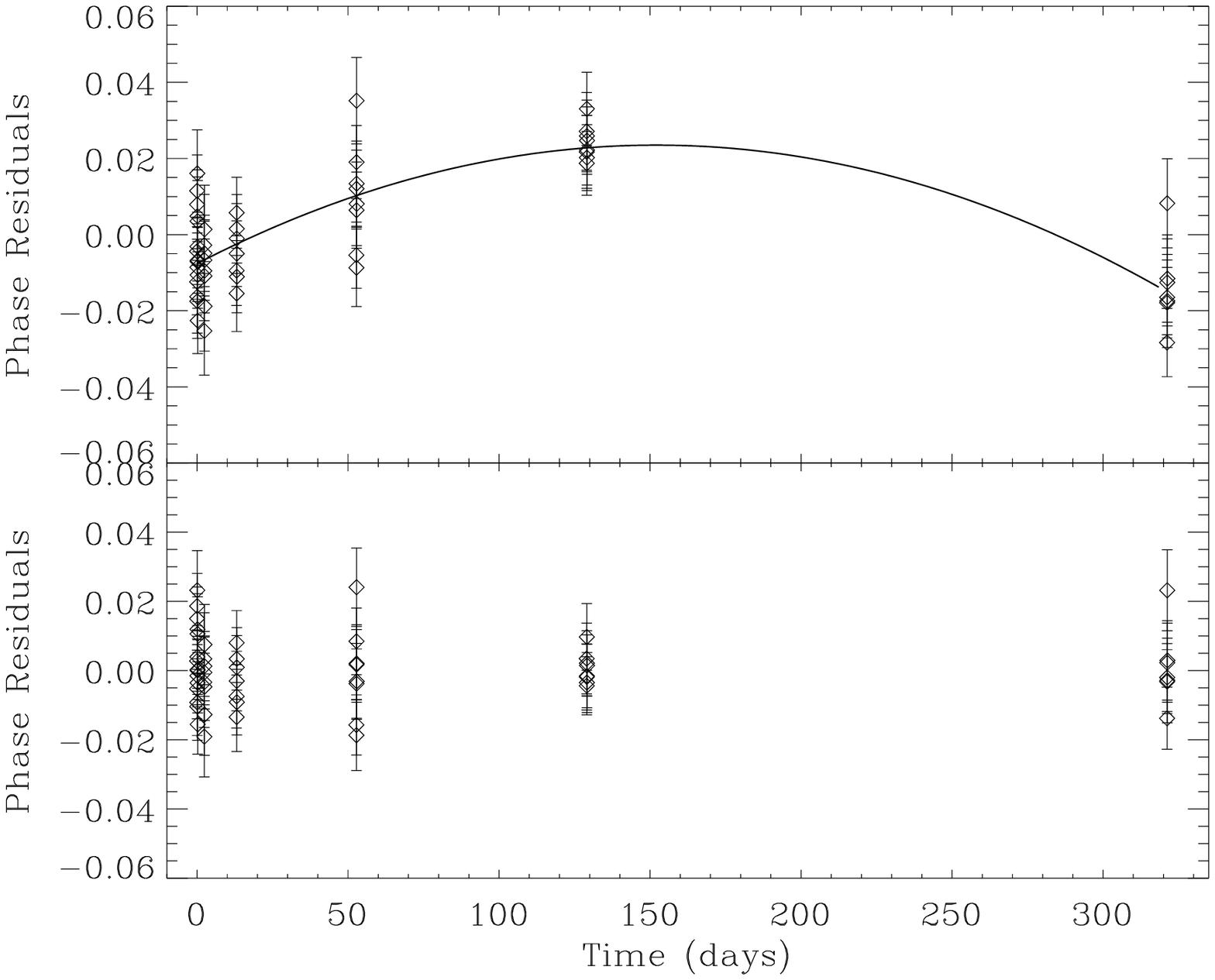]{Phase-timing residuals from our epoch 2004
monitoring observations of J0806. Plotted as a function of time are
the residuals using the best constant frequency phase model (i.e.,
$\dot\nu = 0$; top panel), and with the best solution including a
positive $\dot\nu = 3.77 \times 10^{-16}$ Hz s$^{-1}$ (bottom
panel). In the top panel a quadratic trend with the parabola opening
downward is indicative of the need for a positive $\dot\nu$. The zero
point on the time axis corresponds to $t_0 = 53009.889943753$ MJD
(TDB).
\label{f5}}

\figcaption[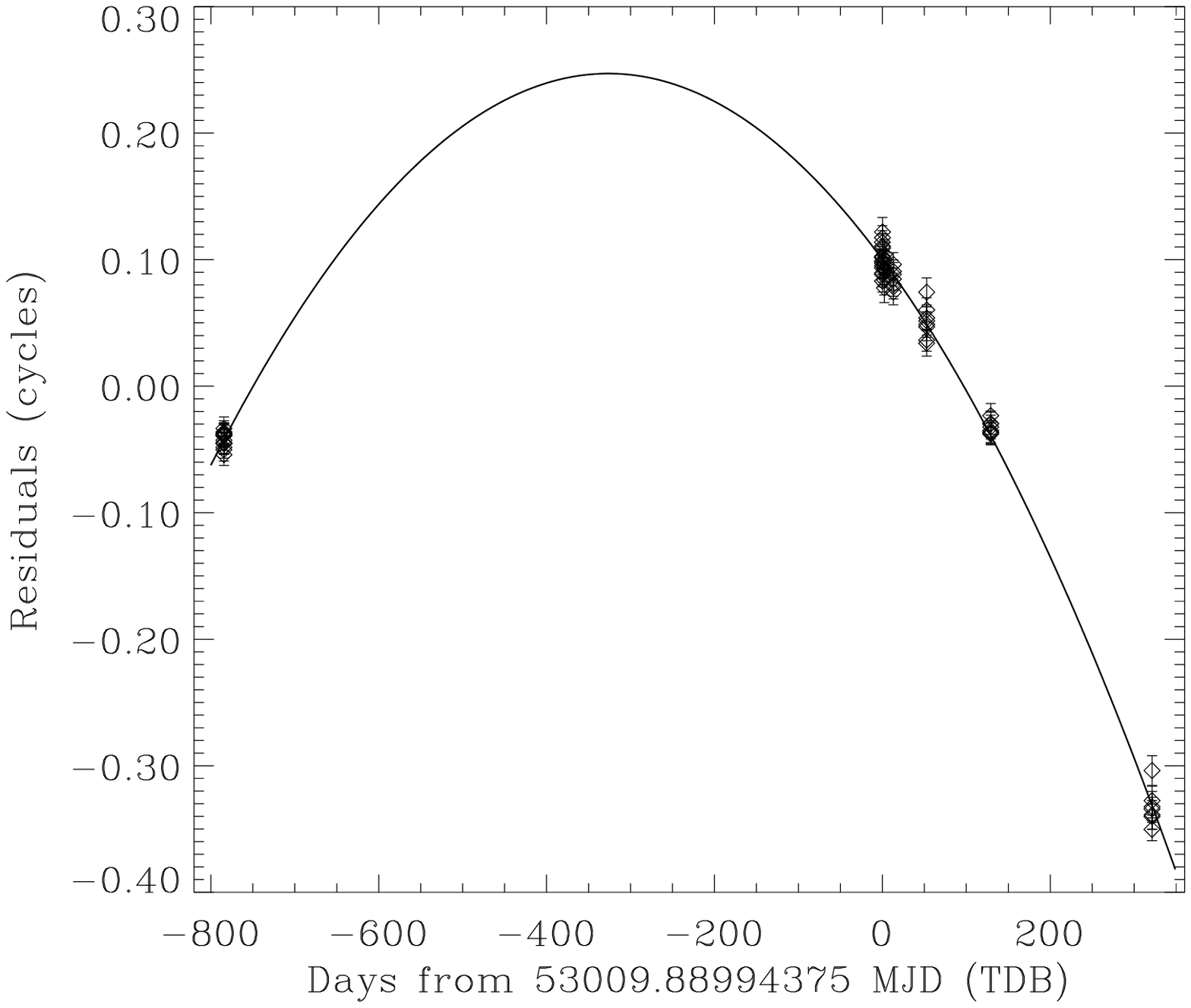]{Phase timing measurements for J0806 including the
November, 2001 (UTC) {\it Chandra} observations.  This plot shows how
the phases drift with respect to a constant frequency model. The solid
curve shows the spin-up model deduced from the 2004 epoch data, and is
consistent with the phase timing of the epoch 2001 data.
\label{f6}}

\figcaption[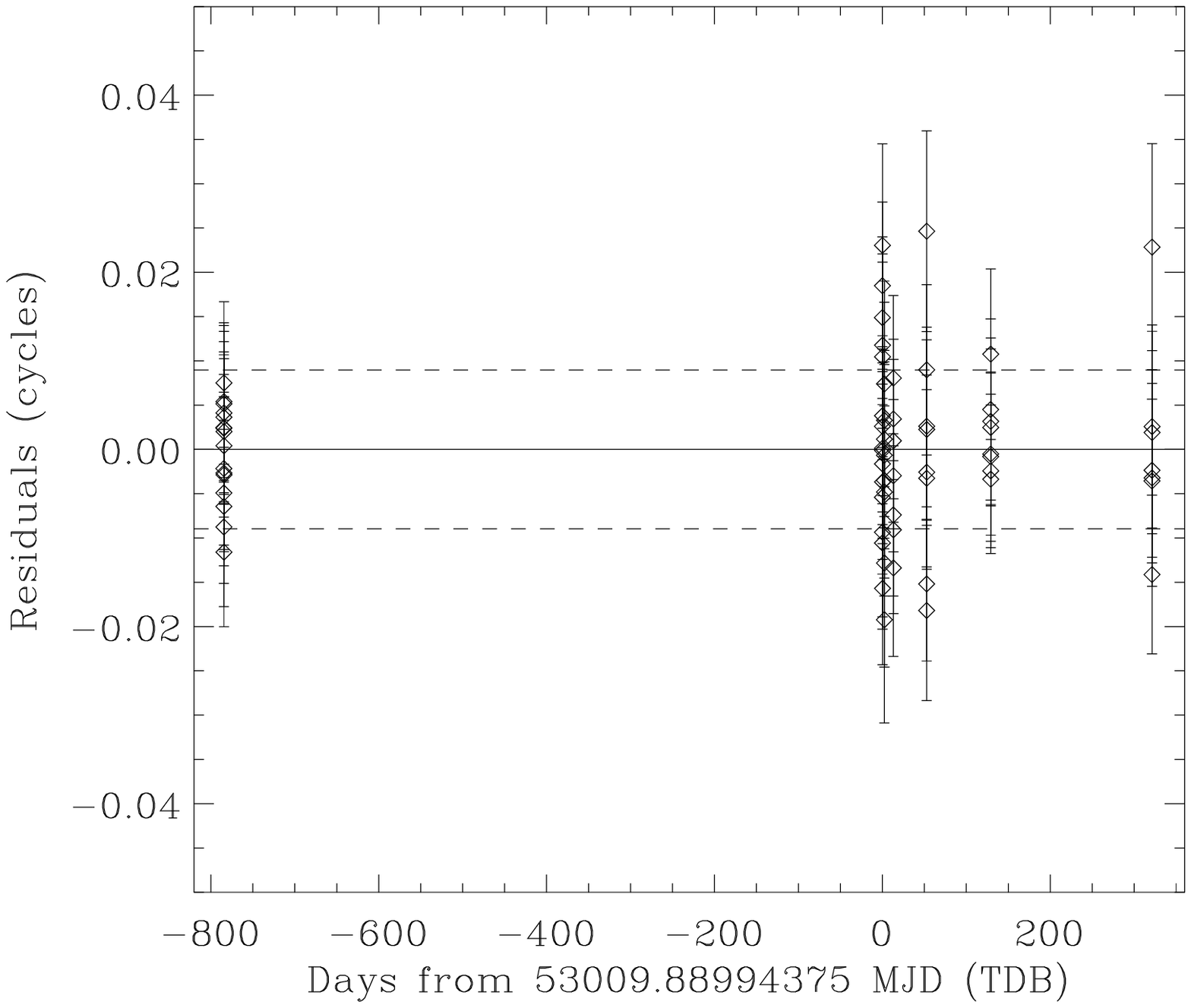]{Same as Figure 6 but now the model includes the
best fit value of $\dot\nu = 3.63 \pm 0.06 \times 10^{-16}$ Hz
s$^{-1}$. The horizontal dashed lines show the rms level of the
residuals, which is less than 1/100 of a cycle.
\label{f7}}

\figcaption[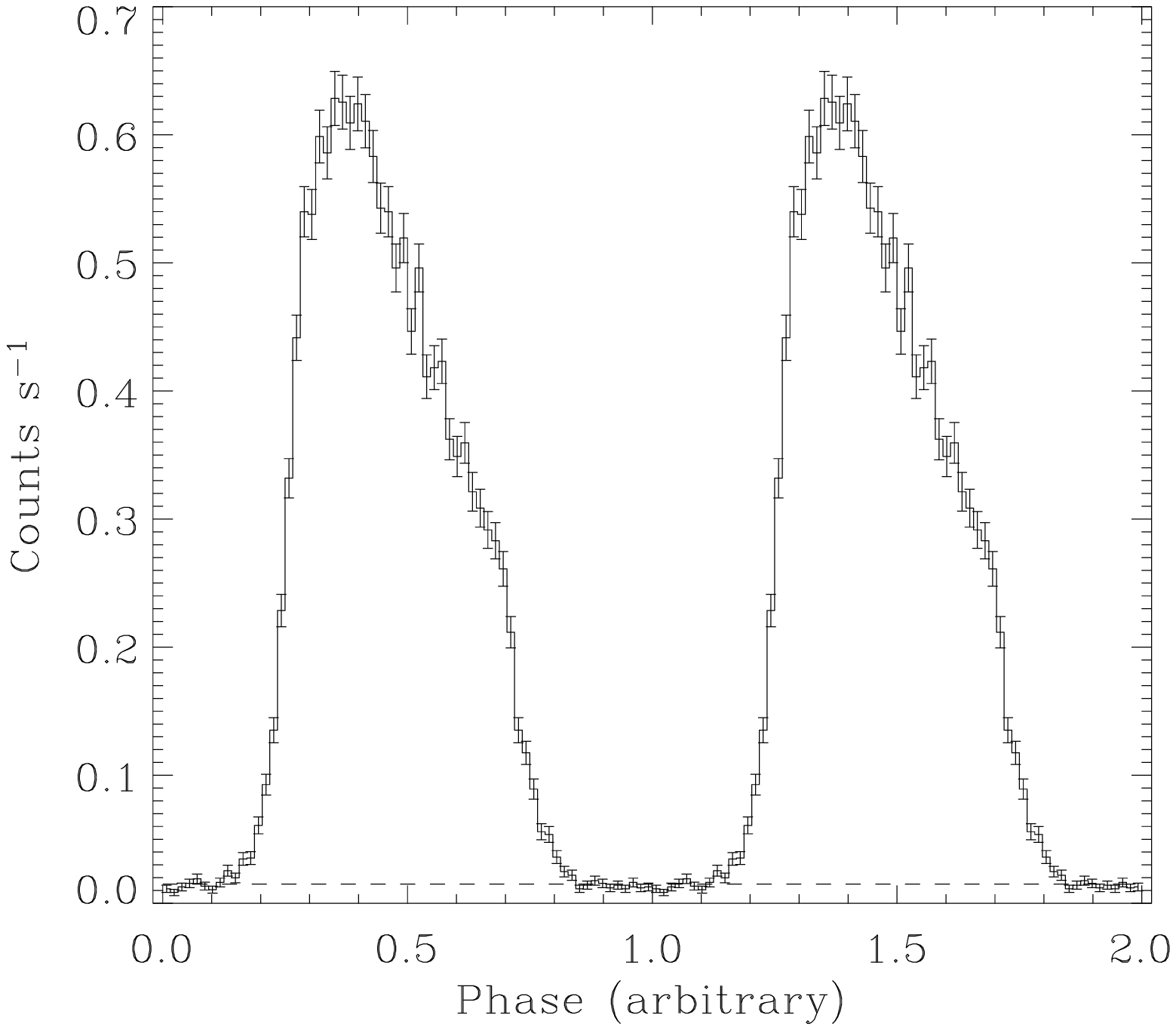]{Folded pulse profile for J0806 using the 2001 and
2004 {\it Chandra} data and our best timing solution. Phase zero is
arbitray, and two cycles are shown for clarity. The horizontal dashed
line is an estimate of the background in the 1-d images associated
with the CC-mode data.
\label{f8}}

\figcaption[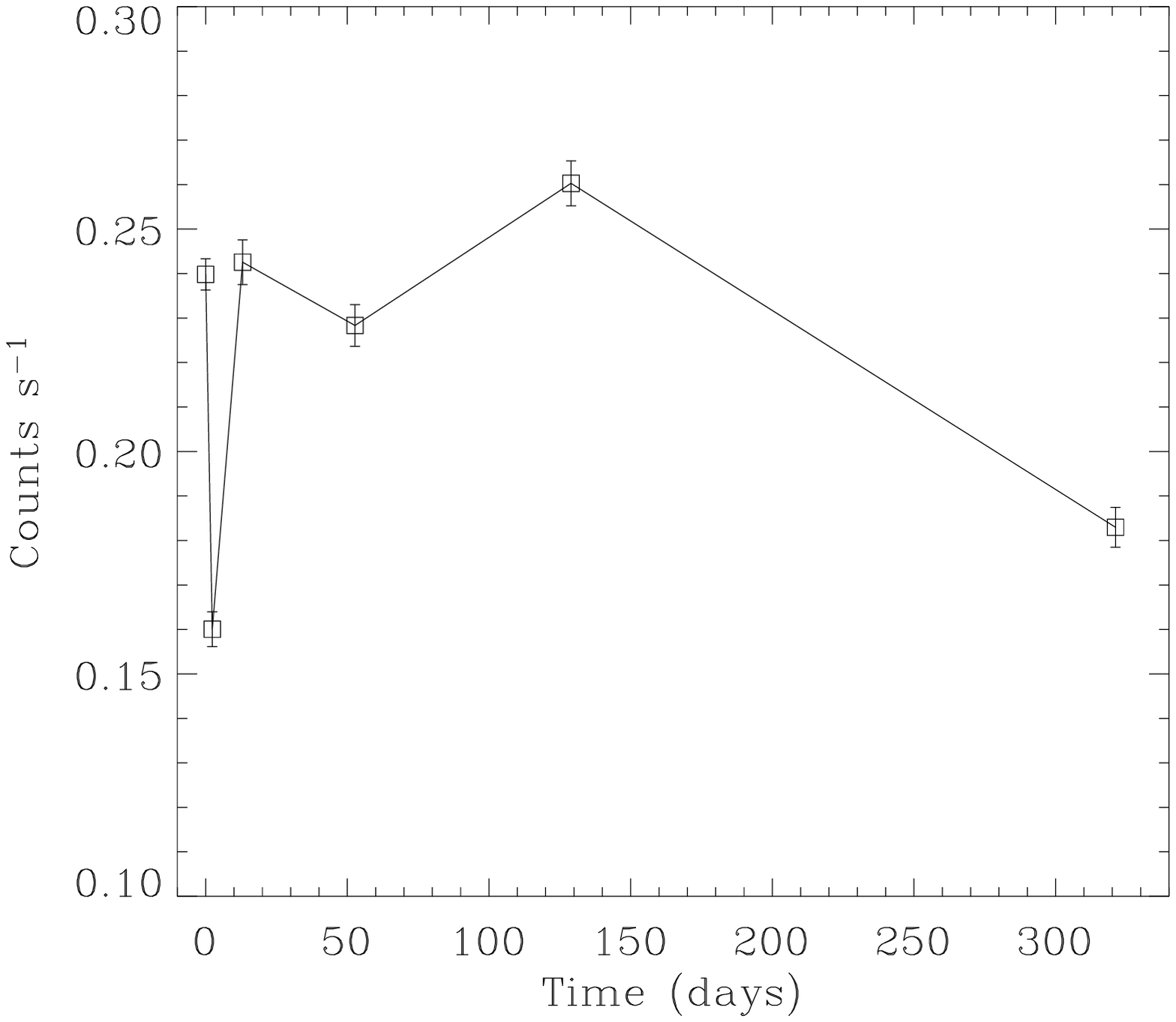]{Average ACIS-S Count rates versus time from J0806
for our 2004 observations. Time is measured from $t_0 =
53009.889943753$ MJD (TDB).  The mean rate is 0.219 s$^{-1}$, and 30\%
variations about this value are evident.
\label{f9}}

\pagebreak

\begin{figure}
\begin{center}
 \includegraphics[width=6in, height=6in]{f1.ps}
\end{center}
Figure 1: {\it Chandra} one-dimensional image of J0806 from our January 5,
2004 (UTC) ACIS-S CC-mode observation. The sharp peak contains photons
from the source.
\end{figure}
\clearpage

\begin{figure}
\begin{center}
 \includegraphics[width=6in, height=6in]{f2.ps}
\end{center}
Figure 2: Light curve of a portion of the {\it Chandra} ACIS-S data from the
January 5, 2004 (UTC) observation of J0806. The bin size is 16 s, and
ten individual pulses are shown.
\end{figure}

\clearpage

\begin{figure}
\begin{center}
 \includegraphics[width=6in, height=6in]{f3.ps}
\end{center}
Figure 3: Best constant frequency measurement for J0806 using all of
our epoch 2004 {\it Chandra} observations.  The $Z_3^2$ power spectrum is
shown as a function of frequency in the vicinity of $\nu_0 =
3.11014250$ mHz.
\end{figure}

\clearpage

\begin{figure}
\begin{center}
 \includegraphics[width=6in,height=6in]{f4.ps}
\end{center}
Figure 4: Constraints on $\nu_0$ and $\dot\nu$ for J0806 from our
phase-timing analysis using the new, epoch 2004 {\it Chandra} dataset
(dashed contours), and after combining the 2001 and 2004 datasets
(solid contours). We show the joint 68\% and 90\% confidence regions
in each case.  The results definitively show that J0806 is spinning up
at at rate of $3.6 \times 10^{-16}$ Hz s$^{-1}$. Here $\nu_0 =
3.1101380$ mHz, and the reference epoch, $t_0 = 53009.889943753$ MJD
(TDB). 
\end{figure}

\clearpage

\begin{figure}
\begin{center}
 \includegraphics[width=6in,height=6in]{f5.ps}
\end{center}
Figure 5: Phase-timing residuals from our epoch 2004 monitoring
observations of J0806. Plotted as a function of time are the residuals
using the best constant frequency phase model (i.e., $\dot\nu = 0$;
top panel), and with the best solution including a positive $\dot\nu =
3.77 \times 10^{-16}$ Hz s$^{-1}$ (bottom panel). In the top panel a
quadratic trend with the parabola opening downward is indicative of
the need for a positive $\dot\nu$. The zero point on the time axis
corresponds to $t_0 = 53009.889943753$ MJD (TDB).
\end{figure}

\clearpage

\begin{figure}
\begin{center}
 \includegraphics[width=6in, height=6in]{f6.ps}
\end{center}
Figure 6: Phase timing measurements for J0806 including the November,
2001 (UTC) {\it Chandra} observations. This plot shows how the phases drift
with respect to a constant frequency model. The solid curve shows the
spin-up model deduced from the 2004 epoch data, and is consistent with
the phase timing of the epoch 2001 data. 
\end{figure}

\clearpage

\begin{figure}
\begin{center}
 \includegraphics[width=6in, height=6in]{f7.ps}
\end{center}
Figure 7: Same as Figure 6 but now the model includes the best fit
value of $\dot\nu = 3.63 \pm 0.06 \times 10^{-16}$ Hz s$^{-1}$. The
horizontal dashed lines show the rms level of the residuals, which is
less than 1/100 of a cycle.
\end{figure}

\clearpage

\begin{figure}
\begin{center}
 \includegraphics[width=6in, height=6in]{f8.ps}
\end{center}
Figure 8: Folded pulse profile for J0806 using the 2001 and 2004
{\it Chandra} data and our best timing solution. Phase zero is arbitray, and
two cycles are shown for clarity. The horizontal dashed line is an
estimate of the background in the 1-d images associated with the
CC-mode data. 
\end{figure}

\clearpage

\begin{figure}
\begin{center}
 \includegraphics[width=6in, height=6in]{f9.ps}
\end{center}
Figure 9: Average ACIS-S Count rates versus time from J0806 for our
2004 observations. Time is measured from $t_0 = 53009.889943753$ MJD
(TDB).  The mean rate is 0.219 s$^{-1}$, and 30\% variations are
evident.
\end{figure}

\clearpage

\begin{table*}
\begin{center}{Table 1: {\it Chandra} Observations of RX J0806.3+1527}
\begin{tabular}{cccccc} \\
\tableline
\tableline
 &  OBSID     &  Instrument    &  Start (TT)   &  Stop (TT)  &  Exp (ksec) \\
\tableline
1 & 300117  &  ACIS-S (CC-mode)  & Jan 05, 2004:21:02:50 & Jan 06, 
2004:02:47:15 & 19.5 \\
2 & 300118  &  ACIS-S (CC-mode)  & Jan 08, 2004:05:59:15 & Jan 08, 
2004:09:30:19 & 10.5 \\
3 & 300119  &  ACIS-S (CC-mode)  & Jan 18, 2004:22:55:48 & Jan 19, 
2004:02:02:29 &  9.8 \\
4 & 300120 &   ACIS-S (CC-mode)  & Feb 27, 2004:15:04:59 & Feb 27, 
2004:18:26:26 & 10.4 \\
5 & 300121 &   ACIS-S (CC-mode)  & May 13, 2004:21:25:04 & May 14, 
2004:00:44:47 & 10.2 \\
6 & 300143 &   ACIS-S (CC-mode)  & Nov 22, 2004:01:36:42 & Nov 22, 
2004:04:32:39 &  9.2 \\
\tableline
\end{tabular}
\end{center}
\end{table*}
\clearpage


\end{document}